# Dielectric Optical-Controlled Magnifying Lens by Nonlinear Negative Refraction

**Jianjun Cao[1], Ce Shang[2], Yuanlin Zheng[1], Xianfeng Chen[1,3], Xiaogan Liang[4] and Wenjie Wan[1,2,3*]**

[1]Key Laboratory for Laser Plasmas (Ministry of Education) and Collaborative Innovation Center of IFSA, Department of Physics and Astronomy, Shanghai Jiao Tong University, Shanghai 200240, China

[2]University of Michigan-Shanghai Jiao Tong University Joint Institute, Shanghai Jiao Tong University, Shanghai 200240, China

[3]The State Key Laboratory of Advanced Optical Communication Systems and Networks, Shanghai Jiao Tong University, Shanghai 200240, China

[4]Department of Mechanical Engineering, University of Michigan, Ann Arbor, Michigan 48109, USA

**A simple optical lens plays an important role for exploring the microscopic world in science and technology by refracting light with tailored spatially varying refractive index. Recent advancements in nanotechnology enable novel lenses, such as, superlens[1,2], hyperlens[3,4], Luneburg lens[5,6], with sub-wavelength resolution capabilities by specially designing materials' refractive indices with meta-materials[7,8] and transformation optics[9,10]. However, these artificially nano/micro engineered lenses usually suffer high losses from metals and are highly demanding in fabrication. Here we experimentally demonstrate for the first time a nonlinear dielectric magnifying lens using negative refraction by degenerate four-wave mixing in a plano-concave glass slide, obtaining magnified images. Moreover, we transform a nonlinear flat lens into a magnifying lens by introducing transformation optics into nonlinear regime, achieving an all-optical controllable lensing effect through nonlinear wave mixing, which may have many potential applications in microscopy and imaging science.**

A traditional optical lens refracts light with a designed spatially varying refractive index to form images; such images can be magnified or de-magnified according to the law of geometrical optics on both surfaces of the lens by linear refraction, e.g. plano-convex lens, plano-concave lens. These images formed by optical lenses have a limited resolution due to the well-known diffraction limit, caused by lack of detections of near-field evanescent waves at the far field [11]. In order to overcome this limit, a slab-like flat lens, namely "superlens", has been demonstrated with sub-diffraction-limited resolution imaging capability in the near field by exploiting the idea of negatively refracted evanescent waves in some carefully engineered meta-materials [1,11,12]. However, images by superlens can only form in the near-field without any magnification. To mitigate these constrains, the concept of hyperlens later was introduced to convert the near-field evanescent waves into propagating ones providing magnification at the far field by the help of transformation optics to enable negative refraction near some hyperbolic dispersion surfaces [3, 13-15].

Besides optics, various forms of these sub-diffraction-limited resolution lenses have recently been realized in many other fields including microwave and acoustic [16, 17]. One major drawback of these lenses is associated with high losses from metallic materials, which are the essential elements bringing in negative permittivity and artificial permeability to enable negative refraction. Meanwhile, fabrications of such nano/micro structures raise additional obstacle for their practical applications.

To address this problem, alternative approaches have been proposed in nonlinear optics to achieve nonlinear version of negative refraction using phase conjugation, time reversal and four wave mixing (4WM) [18-20], where negative refraction can be attained by exploring nonlinear wave mixings with right angle matching schemes. In contrast to those artificially engineered methods i.e. meta-materials and photonic crystals in linear optics, ideally only a thin flat nonlinear slab is required to enable this nonlinear negative refraction [21-26]. Such negative refractions using nonlinear wave mixing have been demonstrated in some thin films with high nonlinearity such as metal and graphite thin film [21, 22]. Moreover, a flat lens utilizing negative refraction of nonlinear mixing waves has successfully shown its 1D/2D imaging ability in our recent work [26]. However, this lens still lacks the magnification capability, which is crucial for imaging applications.

Here we experimentally demonstrate a new type of dielectric magnifying lens based on nonlinear negative refraction by degenerate four-wave mixing with a thin glass slide. A multi-color imaging scheme is realized at millimeter scale by converting original infrared beams into negatively refracted visible ones, spatial refractive index of the lens is carefully designed to ensure the magnification. By doing so, we surprisingly turn a de-magnifying plano-concave lens in linear optics into a magnifying one in nonlinear optics. Moreover, inspired by the transformation optics, we successfully transform a non-magnified nonlinear flat lens into a magnifying one by controlling divergence of pumping beams, effectively creating a magnifying lens controlled by another optical beam for the first time. This new imaging theme may offer new platform for novel microscopy applications.

Negative Refraction can occur in a nonlinear degenerate four-wave mixing scheme [18, 20] as shown in Fig. 1a, where a thin slab of third order nonlinear susceptibility $\chi^{(3)}$ can internally mix an intense normal-incident pump beam at frequency $\omega_1$ with an angled-incident probe beam at frequency $\omega_2$, generating a 4WM wave at frequency $\omega_3 = 2\omega_1 - \omega_2$ which is negatively refracted with respect to the probe's incidence [21, 26]. Such nonlinear negative refraction arises from the momentum requirement of the phase matching condition: $k_3 = 2k_1 - k_2$ during 4WM in order to ensure efficient wavelength conversion. This phase matching condition can be further translated to a Snell-like angle dependence law and create an effective negative refractive index $n_e$ as (see supplement and Ref.[26]):

$$n_e = \frac{\sin\theta_2}{\sin\theta_3} = -\frac{\lambda_2}{\lambda_3} \qquad (1)$$

where the ratio of *sine* values between the probe's incident angle $\theta_2$ and the 4WM's refraction angle $\theta_3$ (Fig. 1a) negatively proportions to the ratio between their wavelengths $\lambda_2$ and $\lambda_3$. The negative sign indicates reversed angles with respect to the central pump's axis, effectively creating a "negative refraction" between the probe and the 4WM wave. Meanwhile, the phase matching condition in three-dimensional wave vector space (Fig. 1b) exhibits a double cone shape around the central pump's axis, where the joint points between the incident probe's wave vector $k_2$ and the 4WM's wave vector $k_3$ compose a ring in the transverse plane. Physically, this means all the incident probe beams with angles parallel to $k_2$ emitted from a point source will be negatively refracted through 4WM waves and focus on the other side of the slab. This builds the foundation for imaging using such negative refraction by nonlinear four-wave mixing with a thin nonlinear flat slab in Ref.[26]. Both 1D and 2D images can be obtained by a nonlinear flat lens in such manner. However, due to one-to-one correspondence between object points and image points, the images' sizes are the same as the objects' without any magnification, similar to the case with a superlens[1]. In order to overcome this magnification issue, a negative diverging lens, e.g. plano-concave lens, can be combined with nonlinear negative refraction to reduce the converging angles of 4WM beams such that a real magnified image can be obtained, as shown in Fig.1c. As contrast, such plano-concave lens in linear optics only forms a de-magnified virtual image with the same color; our nonlinear plano-concave lens can magnify the image with another color through nonlinear 4WM.

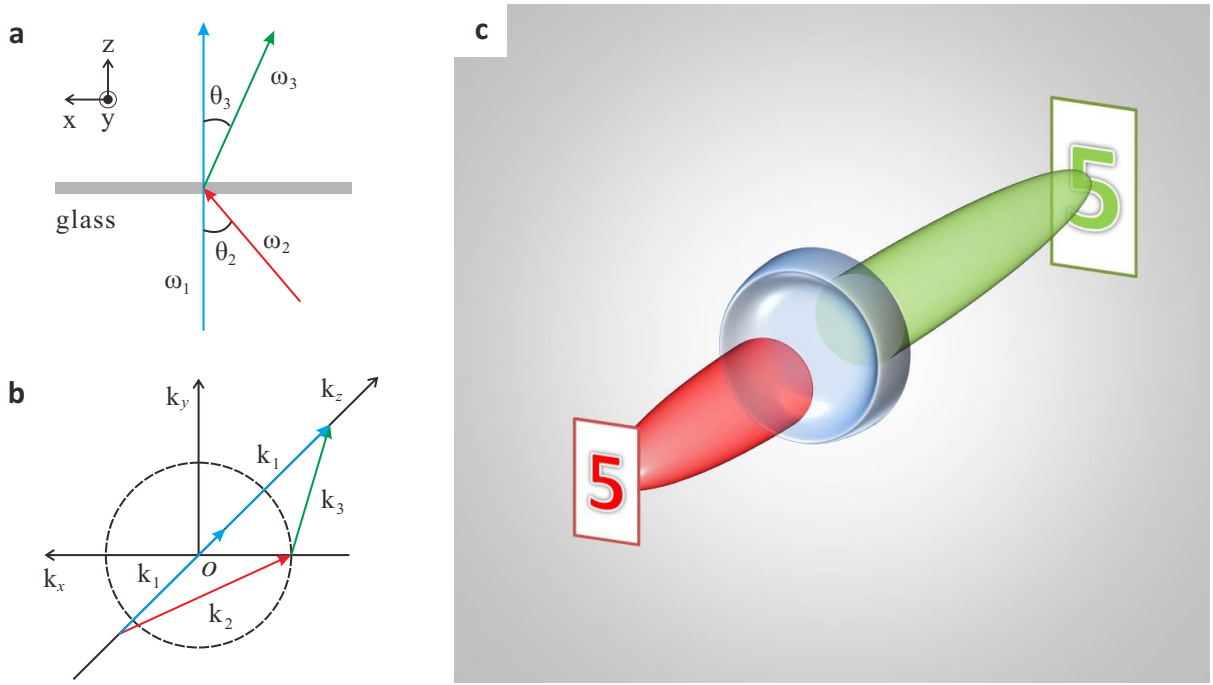

**Figure 1 | Illustration of magnifying lens by nonlinear 4WM. a**, Schematic of negative refraction realized by 4WM process in a thin planar glass slide. In a special case when the pump beam at frequency $\omega_1$ is incident normally on the glass slide, the generate 4WM beam at frequency $\omega_3$ will refract negatively with respect to the angle of the probe beam at frequency $\omega_2$. **b**, The phase matching condition for the degenerate 4WM process in 3D wave vector space. The dashed ring line indicates the joint points of wave vector $k_2$ and $k_3$ that fulfill phase matching condition: $2k_1 - k_2 - k_3 = 0$. **c**, Schematic of the experimental setup of magnifying lens by 4WM. The probe beam $\omega_2$ that carries the object information can nonlinearly mix with the pump beam $\omega_1$ in

a plano-concave lens to give rise to the 4WM beam which can form the magnified image of the object.

To elaborate this idea, we consider a four-wave mixing process near a plano-concave lens as shown in Fig. 2a. An intense normal incident pump beam can nonlinearly mix with a probe beam with an incidence angle matching 4WM phase matching condition in Fig. 1b to generate a 4WM beam. In a nonlinear flat lens (i.e. double plano-surface slab)[26], such 4WM beams can be negatively refracted with respect to the probe as shown as the dash lines in Fig.2a according to the nonlinear refraction law in Equ. (1). With a plano-concave lens, this nonlinear negative refraction can be weakened by the linear Snell's refraction law on the concave surface (solid green lines in Fig.2a), giving a magnified image. Therefore, by combining both the nonlinear refraction law and the linear Snell's law, we can obtain the magnification as (see supplement):

$$M_1 = a \cdot \frac{v}{u} = \frac{1}{1-\frac{u}{f} \cdot a} \qquad (2)$$

where $a = \frac{tan\theta_2}{tan\theta_3}$, $\theta_2$ and $\theta_3$ are the probe's incident angle and 4WM's refraction angle. $f$ is the focal length of the plano-concave lens and $u$ and $v$ are the object distance and the image distance from the lens.

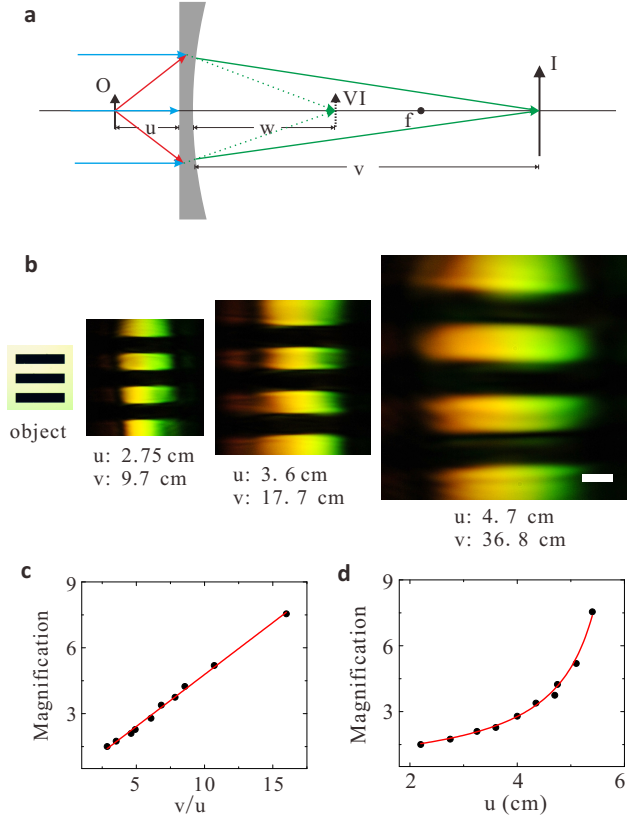

**Figure 2 | Imaging law of the nonlinear magnifying lens using negative refraction. a**, Schematic of the imaging behavior of the magnifying lens. "O", "VI" and "I" stand for object, virtual image and image respectively. "u, w, v, f" are object distance, virtual image distance, image distance and focal length. "VI" represents a virtual image formed by a thin nonlinear flat lens. **b**, Experimental captured images with different magnifications by varying object distances "u". The images are recorded at image distances "v" where they are clearest. The scare bar is 500 μm. **c, d**, The magnification factor as a function of the ratio of the negative image distance and the object distance "-v/u" and the negative object distance "-u". The black circles are experimentally measured data. Solid red lines in **c** and **d** are theoretical curve according to Equ. (2).

In our experiment, the pump beam with the pulse duration of ~75 fs and central wavelength $\lambda_1 = 800$ nm is delivered by a Ti:Sapphire femtosecond laser source while another optical parametric amplifier provides pulses of similar duration at wavelength $\lambda_2 = 1300$ nm as the probe beam. A plano-concave lens made of BK-7 glass with focal length $f = -13.5$ mm, is used as our nonlinear lens which contains the third order nonlinear susceptibility $\chi^{(3)}$ around $2.8 \times 10^{-22}$ m²/V² [27]. The incident angle of the probe beam $\theta_2$ is set to 7.4°, close to the phase matching condition inside BK-7 glass material in order to ensure nonlinear wave conversion about $10^{-5}$ efficiency. In a non-collinear configuration, a USAF resolution card is placed on the probe's path with a distance $u$ away from the lens, while the images formed with 4WM beams around 578nm wavelength can be captured by a CCD camera. Fig. 2b shows such images with different magnifications by varying the object distance. The measured magnification linearly proportions to the ratio between the image and object distances as shown in Fig. 2c: the linear fitting slope reads 0.473, similar to 0.468 calculated from Equ. (2). Fig. 2d further proves the validity of Equ. (2) by only varying the objective distance $u$, showing good agreement between experimental measurements and the theory. It is also worth mentioning that the rainbow colors in the images result from multicolor 4WM processes enabled by the slight phase mismatching inside nonlinear glass due to finite spectrum spreading of the incoming beams. Such colors can also be affected by the glass slide's thickness as discussed previously in Ref.[26], also in supplement.

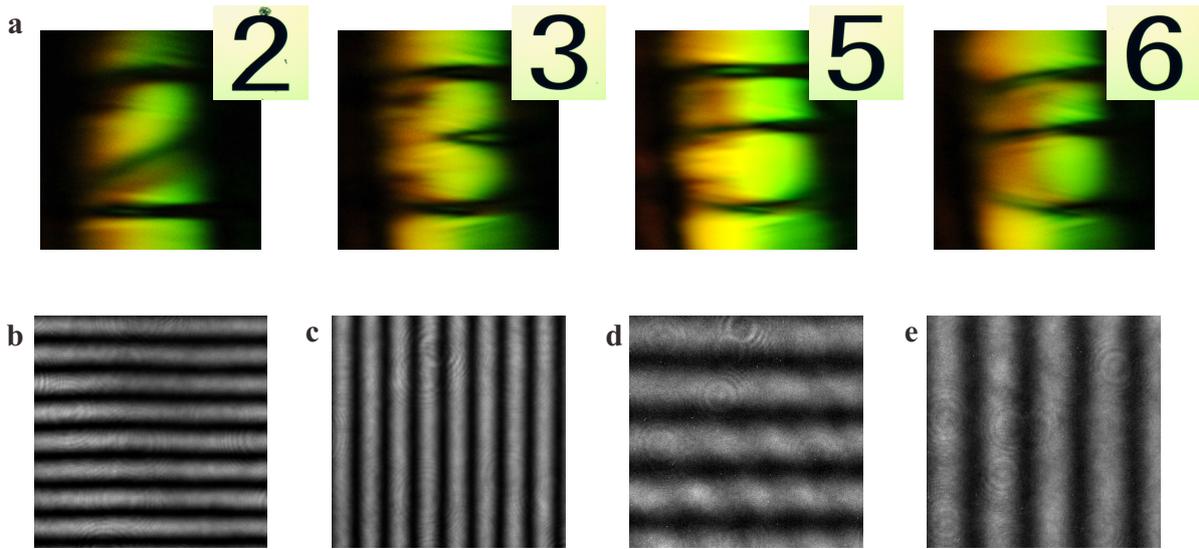

**Figure 3 | Experimental 2D images formed by the magnifying lens in non-collinear and collinear configurations**. **a**, Magnified images of the "numbers" in the USAF resolution card recorded at u = −3.25 cm, v = 15 cm in a non-collinear experimental setup. The corresponding original object images are shown in the insets. The scare bar is 500 μm. **b-e**, Images of the gratings in a collinear experimental setup. **b**, Object image with horizontal lines. **c**, Object image with vertical lines. **d**, Magnified image of the object with horizontal lines. **e**, magnified image of the object with vertical lines. The scare bar is 10μm.

Figure 3a shows 2D magnified images formed by the nonlinear magnifying lens in the non-collinear configuration. It is noticeable that the horizontal features are much clearer than the vertical ones. This is because the incident pump and probe beams both lay on the same horizontal plane, where only one small portion of phase matching ring near the horizontal plane in Fig. 1b is exploited, giving a better phase matching to 4WMs on that plane, while not to 4WMs on the vertical one (see supplement and Ref.[26] for more details). Hence, 4WMs can be better generated and focused near the horizontal plane, giving a finer resolution. To overcome this limitation, we implement a collinear configuration to access the full phase matching ring in 3D vector space in Fig. 1b (details in supplement), where a normal incident pump beam combined with probe beams scattered off the image object can fulfill the phase matching around the full ring geometry in 3D vector space (Fig. 1b) to generate 4WMs. Unlike the non-collinear configuration, both vertical and horizontal lines are clear now in Fig. 3d, e with a magnification around ~1.87 given by Equ. (2).

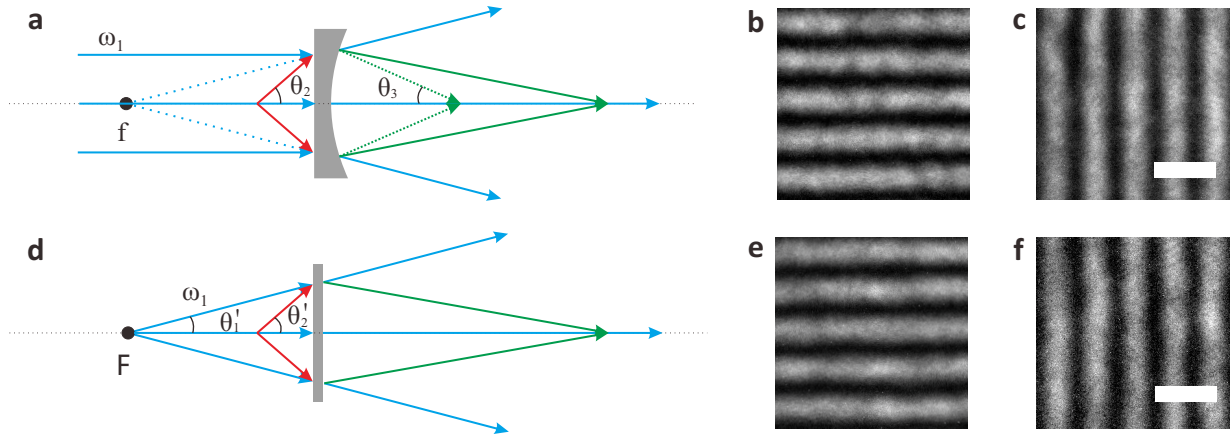

**Figure 4 | Transforming a flat lens into a magnifying lens. a**, Schematic of a nonlinear plano-concave magnifying lens: normally incident pump beams are diverged by the lens. "f" is the virtual focus of the plano-concave lens in linear optics. **b,c**, Magnified images of the gratings formed by the nonlinear plano-concave lens with focal length f=-13.5cm. **d**, Schematic of a nonlinear magnifying flat lens: the pump beam emits from the point "F", diverged along the same paths as the former case behind the flat lens. 4WMs can be generated in a similar manner in both cases. **e,f**, Magnified images of the gratings formed by a flat lens with a diverged pump beam 13.5cm away from the lens. The scare bar is 10 μm.

Inspired by the development of transformation optics [9, 10], we can transform a non-magnifying nonlinear flat lens[26] into a magnifying one by connecting the spatially varying index in a plano-concave nonlinear magnifying lens to the 4WM phase match conditions (effective negative refractive index ) in a non-magnifying nonlinear flat lens. Fig. 4 illustrates this idea: with a nonlinear plano-concave magnifying lens mentioned above, the pump beam usually is normally incident to the front facet of the lens, diverged by the plano-concave lens due to linear refraction (Fig. 4a). This behavior can be mimicked by a point-like divergent pump beam passing through a flat slide (Fig. 4d). Meanwhile, 4WMs in Fig. 4d no longer fulfill the same phase matching uniformly along the transverse plane as in Fig. 1b due to the spatially varying incidence of the divergent pump beam, effectively experiencing spatially varying negative refractive index [21, 26] similar to the linear case of light propagation inside a gradient index (GRIN) lens transformed from a plano-convex lens. While traditional transformation optics relies on artificial meta-materials to produce spatial variations to manipulate the light propagation in a linear fashion, our method here creates the first example ever using effective negative refractive index by nonlinear 4WMs. By considering both nonlinear 4WMs and linear refraction of concave surfaces, we can further derive the magnification factors as below within a paraxial approximation owing to the relative small phase matching angle ~7.6° between the probe and the pump, see supplement:

$$M_2 = \frac{1}{1-\frac{u}{F}\frac{\tan\theta_2'}{\tan\theta_3}} \quad (3)$$

where $M_2$ are the magnifications of a nonlinear flat lens with a divergent pumping. In Fig. 4d, $\theta_2'$ is the probe's incident angle. $F$ is the distance between the pump and the lens. Technically, $\theta_2'$ is different from $\theta_2$ in Equ.(2) which is the probe's incident angle in a nonlinear plano-concave lens in Fig. 4a because they have to fulfill different phase matching due to the pump's incidence. In our case, this difference is only ~0.7°, which is within the allowed 4WM angle spreading due to multicolor spectrums of pump/probe beams and lens's thickness effect during 4WMs explained in supplement and Ref.[26]. This makes the equivalence of Equ. (2) and Equ. (3) if $f = F$. Experimentally, we confirm this by transforming a nonlinear plano-concave lens with $f = -13.5\text{cm}$ to a nonlinear flat lens with a divergent pump 13.5cm away from the lens, obtaining the 2D images with similar magnification ~1.26 in both cases as shown in Fig. 4 b,c,e,f with a collinear configuration (details in supplement).

At last, we show the most interesting feature by this transformed nonlinear lens: optical controlled magnification. Note that compared to Equ.(2), Equ. (3) contains the effective focal length $F$, which can be tunable by tuning the divergence point of the pump beam, effectively optically controlling the nonlinear lens' focus length. By varying this effective focus, we can control the magnification of formed images. For example, we experimentally can increase the magnification to 1.58 from 1.31 in Fig. 5b,c,e,f by decreasing $F$ from -10cm to -6cm. This create the first example ever of an optical controllable lens, as all previous works involves mostly with liquid crystal, thermal effect or deformed liquid lenses[28-30], which could have slow responsibility. Such optical controllable devices may trigger new applications in imaging science.

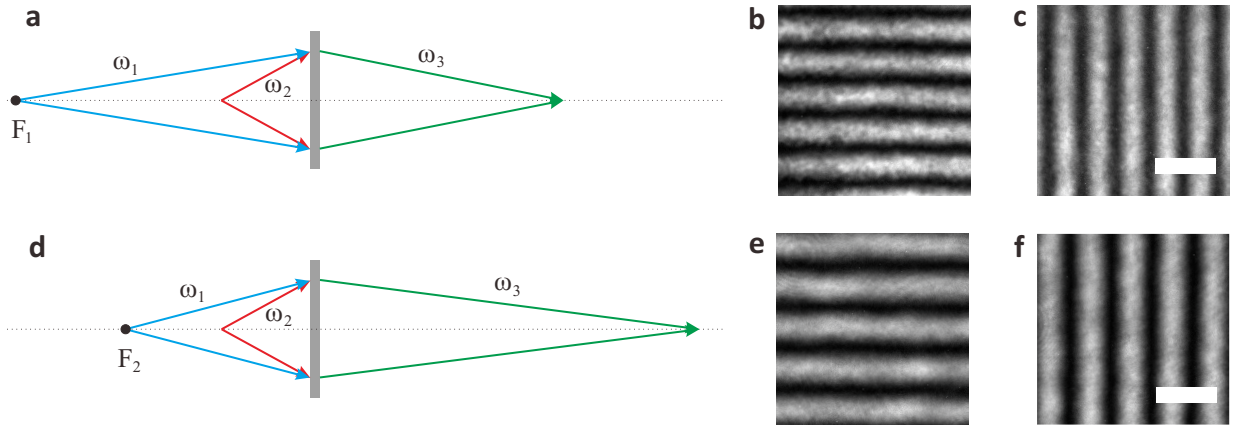

**Figure 5 | Optical controlling a nonlinear magnifying flat lens. a,d**, Schematic of a nonlinear magnifying flat lens with the pump distance $f_1 = 10$ cm, $f_2 = 6$ cm. **b,c**, Magnified images of the gratings formed by the nonlinear magnifying flat lens in **a** with magnification 1.31. **e,f**, Magnified images of the gratings formed by the nonlinear magnifying flat lens in **d** with magnification 1.58. The scare bar is 10 μm.

In conclusion, we have experimentally demonstrated a dielectric nonlinear magnifying lens by nonlinear refraction through four-wave mixing in a thin glass slide. Our method explores the possibility of using dielectric's nonlinear properties for negative refraction as a substitute approach for meta-materials to overcome the loss problem. We extend the transformation optics into nonlinear regime for the first time, creating a nonlinear optical-controlled magnifying lens. The new nonlinear optical lens design reported here may open new realms of many applications in microscopy and imaging science in the near future.

**Acknowledgements**
This work was supported by the National Natural Science Foundation of China (Grant No. 11304201, No. 61125503, No. 61475100), the National 1000-plan Program (Youth), Shanghai Pujiang Talent Program (Grant No. 12PJ1404700), Shanghai Scientific Innovation Program (Grant No. 14JC1402900)

**Correspondence**   requests for materials should be addressed to Wenjie Wan (email: wenjie.wan@sjtu.edu.cn)